\documentclass[12pt,aps,prd,preprint,tightenlines,showpacs,nofootinbib]{revtex4}
\newcommand{\PRE}[1]{{#1}} 

\usepackage{hyperref}      
\usepackage{bm}
\usepackage{epsfig}

\newcommand{\eqref}[1]{Eq.~(\ref{#1})}

\newcommand{\secref}[1]{Sec.~\ref{sec:#1}}

\newcommand{\be}{\begin{equation}}
\newcommand{\ee}{\end{equation}}
\newcommand{\bea}{\begin{eqnarray}}
\newcommand{\eea}{\end{eqnarray}}
\newcommand{\beq}{\begin{equation}}
\newcommand{\eeq}{\end{equation}}
\newcommand{\beqa}{\begin{eqnarray}}
\newcommand{\eeqa}{\end{eqnarray}}

\newcommand{\sz}{\sim0}

\begin{document}

\title{ \PRE{\vspace*{1.5in}} Flavored Gauge-Mediation\\
\PRE{\vspace*{0.3in}} }

\author{Yael Shadmi
}
\affiliation{Physics Department, Technion-Israel Institute of
Technology, Haifa 32000, Israel
\PRE{\vspace*{.5in}}
}
\author{Peter Z. Szabo
}
\affiliation{Physics Department, Technion-Israel Institute of
Technology, Haifa 32000, Israel
\PRE{\vspace*{.5in}}
}

\begin{abstract}
\PRE{\vspace*{.3in}} 
The messengers of Gauge-Mediation Models can couple to standard-model matter
fields through renormalizable superpotential couplings.
These matter-messenger couplings generate generation-dependent
sfermion masses and are therefore usually forbidden by discrete
symmetries. However, the non-trivial structure of the standard-model Yukawa
couplings hints at some underlying flavor theory, which
would necessarily control the sizes of the matter-messenger
couplings as well. Thus for example, if the doublet messenger and the Higgs
have the same properties under the flavor theory, the resulting 
messenger-lepton couplings
are parametrically of the same order as the lepton Yukawas,
so that slepton mass-splittings are similar to those of 
minimally-flavor-violating models and therefore satisfy bounds on 
flavor-violation,
with, however, slepton mixings that are potentially large.
Assuming that fermion masses are explained by a flavor symmetry,
we construct viable and natural models with messenger-lepton
couplings controlled by the flavor symmetry.
The resulting slepton spectra are unusual and interesting, with
slepton mass-splittings and mixings that may be probed at the LHC.
In particular, since the new contributions are typically negative, 
and since they are often larger for the first- and second-generation 
sleptons, some of  these examples have the selectron 
or the smuon as the lightest slepton,
with mass splittings of a few to tens of GeV.
\end{abstract}


\maketitle

\section{Introduction}
\label{sec:introduction}
Motivated by the absence of flavor changing neutral currents
and rare decays,
most studies of supersymmetry at colliders assume  
universal sfermion masses at the scale where supersymmetry breaking
is mediated to the Minimal Supersymmetric Standard Model (MSSM). 
Any sfermion mass splittings or mixings then
originate from the Standard Model (SM)  Yukawa couplings only, 
and are negligibly small,
except for the stop mass splitting and,
for large $\tan\beta$, also the sbottom and stau
mass splittings. Such models, in which the SM Yukawas are the only 
source of generation-dependence, are usually referred to as 
Minimally Flavor Violating (MFV).
The assumption of MFV is too restrictive however. 
Current constraints
on lepton-violating decays~\cite{Brooks:1999pu,Aubert:2009tk} 
for example 
allow for slepton mass splittings
and mixings that may well be observable at the LHC (see for 
example~\cite{flns}).
If such splittings and mixings are indeed observed, they would
provide a wealth of information about the origin of supersymmetry
breaking, and quite possibly, about the origin of the SM
fermion masses.

It is interesting to ask therefore if there are viable models of supersymmetry
breaking that give rise to appreciable departures from sfermion
mass universality, and several classes of models were recently discussed
in the literature~\cite{flns,Kribs:2007ac,Nomura:2008gg,Nomura:2008pt,
Gross:2011gj}. 
Here we will present another example which is
particularly simple, namely, Minimal Gauge Mediated Supersymmetry Breaking
(GMSB) Models~\cite{Dine:1994vc,Dine:1995ag} with
messenger-matter couplings.

The main appeal of GMSB models of course is that the soft masses are
generated by gauge interactions and are therefore generation-independent
by construction.
In practice, however, in the most successful examples of GMSB, the soft masses
are generated by loops of messenger fields with SM gauge quantum numbers,
which can have renormalizable superpotential couplings
to the MSSM~\cite{Dvali:1996cu,dns,gr,CP}.  
Such couplings would lead to generation-dependent sfermion 
masses and flavor changing neutral currents,
so one usually invokes some global symmetries in order to forbid them. 
Here we will take a more liberal approach towards
messenger-matter couplings, and show that they can result in viable models
with rich and interesting spectra\footnote{Matter-messenger couplings
were also studied in the context of triplet seesaw models,
with the messengers in a 15$+\bar{15}$ of 
SU(5)~\cite{Joaquim:2006uz,Joaquim:2006mn,Brignole:2010nh}.}.

Consider for example the standard set of vectorlike $5+\bar5$ 
messengers~\cite{Dine:1994vc,Dine:1995ag}.  
Of these, one of the SU(2) doublets, which we will denote by $D$,
is in the same SM representation as the down-type Higgs, $H_D$.
Assuming that it has the same R-parity assignment as the Higgs,
the superpotential can contain terms of the form
\beq\label{basic}
y_L D\, l\, e\ ,
\eeq
in addition to the usual Yukawa
\beq
Y_L H_D\, l\, e\ .
\eeq
Here $l$ is the lepton doublet, $e$ is the lepton singlet, and $Y_L$ and $y_L$
are $3\times3$ matrices of couplings.
For an arbitrary matrix $y_L$, one would have disastrous flavor 
changing processes. 
However, the SM Yukawa matrix $Y_L$ is far from
arbitrary. Most of its entries are very small, hinting at some underlying
flavor theory. So it is not implausible that the same underlying
theory would also suppress the entries of the new coupling $y_L$,
so that the two matrices are of the same order of magnitude.
All flavor constraints would then be satisfied, because the model
is qualitatively MFV: all generation dependence originates from
couplings which, albeit new, are of the same order of magnitude as the SM
Yukawas, and the resulting mass splittings are similar
to those of MFV models.
Even this minimal scenario has interesting phenomenological
implications. Since the two matrices $Y_L$ and $y_L$ are not
proportional to each other, slepton mixings can be appreciable.
As a concrete example, assume that the  Yukawa matrix is governed by 
an abelian flavor (Froggatt-Nielsen)~\cite{Froggatt:1978nt}
 symmetry. 
If $D$ and $H_D$ carry the same
charge under the flavor symmetry, each entry of the matrix $y_L$ is 
parametrically the same as the corresponding entry in the matrix $Y_L$,
realizing the minimal scenario described above.
As we will see, one can
also construct models in which $y_L$ is very different from $Y_L$,
leading to large splittings between the first two generations, and with the 
selectron or smuon being lighter than the stau.

All our models are GMSB models with matter-messenger couplings controlled
by the same flavor symmetry  which generates the structure
of fermion mass matrices.
Since the slepton masses, and in particular, the selectron and smuon
masses, will probably be the easiest probes of flavor dependence at the 
LHC,
we focus on models in which the only new messenger couplings involve
leptons. The LHC signatures of generation dependent slepton
spectra have received a lot of attention 
recently (see for example~\cite{ArkaniHamed:1996au,ArkaniHamed:1997km,
Agashe:1999bm,Hisano:2002iy,Hisano:2008ng,Kitano:2008en,
Feng:2009yq,Feng:2009bd,Feng:2009bs,Allanach:2008ib,Kaneko:2008re,
DeSimone:2009ws,Buras:2009sg,Ito:2009xy,Fok:2010vk,
Dreiner:2011wm}).
It would be interesting to generalize our results to the 
squarks as well.

In fact,
the largest couplings are often the couplings to the first generations,
so the selectron or smuon exhibit the largest mass splitting.
The reason is that, in order to obtain appreciable 
mass splittings, we need some entries of $y_L$
to be larger than the corresponding entry in $Y_L$.
This happens if the flavor charge of $D$ is smaller than
the flavor charge of $H_D$ (adopting the standard convention
that the flavor spurion has negative charge).
On the other hand, the third-generation fields must
have smaller flavor charges than the first- and second-generation
fields, in order for their masses to be less suppressed.
Generically then, the entries of $y_L$ corresponding to the
third generation have overall negative charge, and 
since the  superpotential can only contain positive powers
of the spurion, cannot appear in the superpotential.

The couplings of~\eqref{basic} were studied in~\cite{CP}, 
motivated by the fact that they mediate messenger decay, 
and thus solve the cosmological problems associated with stable
messengers. 
Unlike our models, the models of~\cite{CP} were MFV,
with the messengers and SM living on different branes in a 5d setup
so that the couplings~\eqref{basic} originate solely from
Higgs-messenger mixings, with $y=\epsilon Y$ and with $\epsilon$
suppressed by the size of the 
extra-dimension\footnote{This can be achieved in  4d too,
using some broken global symmetry to distinguish between 
$D$ and $H_D$, with the suppression factor 
$\epsilon$ being the relevant spurion.}.

We will classify the different possible messenger-matter couplings
in~\secref{general}, and present the basic superpotential
of our models in~\secref{basic}.
In~\secref{examples} we discuss a few example
models and their spectra.

\section{General Matter-Messenger Couplings}
\label{sec:general}
Minimal GMSB models~\cite{Dine:1994vc,Dine:1995ag} 
involve $N_5$ pairs of vector-like
messengers transforming as  $5+\bar5$'s of SU(5),
coupled to a SM gauge singlet $X$, whose vacuum expectation value (VEV) 
$\langle{X}\rangle\equiv M$ gives
mass to the messengers, and whose $F$-term is non-zero,
leading to supersymmetry-breaking splittings in the messenger
spectrum.
Under the SM gauge group, the messengers transform as
\beq
T_I\sim(3,1)_{-1/3}\ \ \ \bar{T}_I\sim(\bar3,1)_{1/3}\ \ \
D_I\sim(1,2)_{-1/2}\ \ \ \bar{D}_I\sim(1,2)_{1/2} \ ,
\eeq
where $I=1\ldots N_5$.
For ease of notation, we will define in the following
\beq
D\equiv D_1 \ .
\eeq

The possible trilinear superpotential couplings of the messengers to the
SM depend on the messengers R-parity charge assignment.
For R-parity odd messengers, the most general trilinear superpotential
is of the form~\cite{dns,gr}
\beq\label{eq:odd}
W_{\rm {odd}}=   H_D q T + H_D D e^c ,
\eeq 
where $q$ denotes the doublet quarks, and $e^c$ denotes the singlet leptons.
This superpotential breaks baryon- and lepton-number.
For R-parity even messengers one can have~\cite{CP,gr}
\beq\label{eq:even}
W_{\rm {even}}=  y_U\bar{D} q u^c + y_D D q d^c +  y_L D l e^c  ,
\eeq 
where $u^c$ and $d^c$ are the singlet up and down quarks
respectively,  $l$ is the lepton doublet,
and the $y$'s are $3\times3$ matrices of couplings.
(Throughout, we use small letters for matter fields to distinguish them
from the Higgses and messengers, which we denote by capital letters.)
Here we assume that the messengers
have the same R-parity charge assignment as the Higgses,
so that the relevant superpotential is~\eqref{eq:even}.

The couplings in~\eqref{eq:odd}, \eqref{eq:even}  
generate sfermion masses squared
starting at one-loop~\cite{dns,gr}, but the one-loop contributions vanish
at leading order in the supersymmetry breaking, so that
in the limit of small supersymmetry breaking, the dominant
contributions are the two-loop analogs of the usual gauge contributions.
Here we will concentrate on these two-loop contributions.
Unlike in minimal GMSB models, the new couplings also generate
$A$-terms at one-loop. 
The dependence of the soft terms on the matrices $Y$ and $y$ can be inferred 
from a spurion analysis as in~\cite{D'Ambrosio:2002ex},
treating $Y$ and $y$ as spurions of the SM SU(3)$^5$ flavor symmetry.
Since the abelian Froggatt-Nielsen symmetry that 
we will invoke in the following 
only determines the matrices $Y$ and $y$ up to $O(1)$  coefficients,
such a spurion analysis is completely adequate for our purposes.
Still, we will explicitly compute the mixed gauge-Yukawa contributions
to the soft terms\footnote{We derive these using the method of~\cite{gr}, 
generalizing the results of~\cite{CP} to the case of 3-generations,
since we are particularly interested in large couplings
of the messengers to the first and second generation scalars.}. 
As we will see, the gauge-Yukawa contributions will be the dominant
contributions in our models, and knowing their signs will
allow us to determine the hierarchy in the slepton spectrum.

Since we are mainly interested in the implications for the
slepton spectrum, our models are constructed so that $y_U$ always
vanishes and $y_D$ is negligible (or zero).
The slepton masses are then,
\beqa
m^2_{\tilde{l}} &=& \frac{1}{128\pi^2} \Bigg[ 
N_5 \left(\frac{3}{4} g_2^4 + \frac{5}{3}  g_Y^4 \right) {\bf 1}
- \left(
\frac{3}{2} g_2^2 + 6  g_Y^2\right) y_L y_L^\dag +\ldots
\Bigg]\, \left|\frac{F}{M}\right|^2
~,\label{eq:mlsqgen}\\
m^2_{\tilde{e}^c} &=& \frac{1}{128\pi^2} \Bigg[
N_5 \left(\frac{20}{3}  g_Y^4\right) {\bf 1} -
\left(  3 g_2^2 + 12  g_Y^2\right) y_L^\dagger y_L + \ldots
\Bigg] 
\,\left|\frac{F}{M}\right|^2
~.\label{eq:mesqgen}
\eeqa
The first terms in~\eqref{eq:mlsqgen}, \eqref{eq:mesqgen} are the usual
GMSB contributions, which are proportional to the number of
messenger pairs $N_5$.
The remaining terms are new contributions and lead to mass splittings 
and mixings among
the different generations. The latter can be appreciable even
for small $y_L$'s, since the GMSB contribution to the soft mass
is proportional to the identity matrix~\cite{flns}.
The ellipses stand for pure Yukawa terms including terms with four powers
of the matrices $y$, and terms with two powers of $y$ and two powers of $Y$,
such as $y_L y_L^\dagger y_L y_L^\dagger$,  
$y_L y_L^\dagger Y_L Y_L^\dagger+{\rm h.c.}$.
Up to order one coefficients, these terms can be determined by an SU(3)$^5$ 
spurion analysis~\cite{D'Ambrosio:2002ex}.
In all of our models, the pure Yukawa terms are negligible compared to the 
mixed gauge-Yukawa terms, so we can safely ignore 
them\footnote{We thank Anna Rossi for pointing out to us an
error in some of the pure Yukawa terms in an earlier version
of this paper.}. 
The $A$ terms are given by,
\beq
A_{L} = -\frac{1}{16\pi^2} \left[{y_L y_L^\dagger} Y_L + 2 Y_L {y_L^\dagger y_L} 
\right] \frac{F}{M}\ .
\eeq

We note that the structure
of our models is similar to the gauge-gravity hybrid models
of~\cite{flns}, in which the universal contribution is also
gauge-mediated, with a gravity-mediated generation-dependent contribution
which is important for a high messenger scale.
In both frameworks, the size of the non-universal contribution
is controlled by a flavor symmetry, and flavor constraints are
satisfied through the interplay of degeneracy and 
alignment~\cite{Nir:1993mx}.

\section{Basic superpotential}
\label{sec:basic}
In addition to R-parity, we will impose a $Z_3 \otimes Z_2$ symmetry
on the theory, with charges given in Table~\ref{tab:new_sym_charges}.
\begin{table}[h]
\centering
\renewcommand{\arraystretch}{1.25}
\begin{tabular}{|c|c|c|c|}
\hline
Superfield & $R$-parity & $Z_3$ & $Z_2$ \\
\hline\hline
$X$ & even & $1$ & even \\ \cline{1-4}
$T_1$ & even & $0$ & odd \\
$\bar{T}_1$ & even & $-1$ & odd \\
$D$ & even & $0$ & even \\
$\bar{D}_1$ & even & $-1$ & even \\
$T_I, \bar{T}_I, D_I, \bar{D}_I~(I = 2,...,N_5)$ & even & $1$ & even \\ 
\cline{1-4}
$q,u^c,d^c,l,e^c$ & odd & $0$ & even \\
$H_U,H_D$ & even & $0$ & even \\
\hline
\end{tabular}
\caption{$Z_3\times Z_2$ symmetry charges.}
\label{tab:new_sym_charges}
\end{table}
The most general superpotential allowed by this symmetry is
\beqa
\label{eq:superpot1b}
W &=& X \left( X X + T_I \bar{T}_I + D_I \bar{D}_I + H_D \bar{D}_1 \right)
+ H_U q u^c + H_D q d^c + H_D l e^c \\ 
&+& D q d^c + D l e^c \nonumber 
~,
\eeqa
where we omitted the generalized $\mu$-terms,  $H_U H_D + H_U D$.
Just like the usual $\mu$-term, these can be forbidden by some Peccei-Quinn
symmetry, and we will not consider them in the following.

The first line of~\eqref{eq:superpot1b} contains the messenger couplings to
the supersymmetry-breaking sector as well as the usual Yukawa terms.
We explicitly display here the term $X^3$, which is 
typically needed in order to generate appropriate VEVs for $X$,
and motivates our choice of a $Z_3$ symmetry. 
We will not consider this term further.

The second line of~\eqref{eq:superpot1b} is our focus here, 
with the messenger field $D$ replacing $H_D$.
The analogous up-type messenger-matter coupling $\bar{D} q u^c$ 
is eliminated by  the $Z_3$ symmetry. 
It is simple to allow for this term as well. To do so,
one must use at least two separate pairs of messengers,
the first charged as shown in Table~\ref{tab:new_sym_charges} for $I = 1$, 
and the second, with the charges of 
$D$ and $\bar{D}$ swapped. 
Since we are interested in slepton masses here, it is simplest
to stick to the charges of Table~\ref{tab:new_sym_charges},
so that the new couplings only involve the leptons and down-quarks.
As we will see later, it is often possible to impose additional symmetries
on the models so that down-quark couplings are eliminated as well.

Note that, in this construction,  the new couplings of the messengers 
to down quarks and leptons (or alternatively, to up-quarks) 
can appear with one set of messengers, $N_5=1$. 
Having both up-type and down-type messenger couplings requires however $N_5>1$. 

\subsection{MFV-like masses} \label{sec:mfvlike}
So far, $H_D$ and $D$ have the same charges under all the symmetries
of the model. If this remains true in the presence of any additional symmetries,
we can define $D$ as the combination of $D$ and $H_D$ that couples to $X$,
and take $H_D$ to be the orthogonal combination.
The superpotential~\eqref{eq:superpot1b} then takes the form
 \beqa
\label{eq:superpot}
W = X \left( X X + T_I \bar{T}_I + D_I \bar{D}_I  \right)
+ Y_U H_U q u^c &+& Y_D H_D q d^c + Y_L H_D l e^c \nonumber\\ 
&+& y_D D q d^c + y_L D l e^c 
~,
\eeqa
where we display also the $3\times3$ matrices of couplings,
with $Y_U$, $Y_D$ and $Y_L$ denoting the usual up-, down-, and lepton-Yukawas
respectively, and $y_D$ and $y_L$ denoting the corresponding new couplings.
In this case, if the Yuakwa matrices are controlled by some underlying
theory, then the matrices $y_L$ and $Y_L$ (and similarly, $y_D$ and $Y_D$) 
are parametrically the same.
These models are therefore quite similar to MFV models. They contain
new matrices of couplings, which, while not proportional to the Yukawa
matrices, satisfy 
\beq\label{eq:mfvlike}
{\left(y_L\right)}_{ij}= c_{ij}\, {\left(Y_L\right)}_{ij} \ ,
\eeq
where $c_{ij}$ are
order-one coefficients and $i,j=1,2,3$ are generation indices.
The resulting mass-splittings are therefore of the same order of magnitude
as those obtained in MFV models. In particular, the first and second
generation scalars are practically degenerate.
As we will see in~\secref{modela}, such slepton mass splittings are 
consistent with bounds on rare-decays even for large mixings.
On the other hand, the inter-generational mixings are model
dependent, and can be large.
The masses of down squarks are more stringently constrained
by bounds on flavor-changing processes, but still,
at least for small $\tan\beta$,
the resulting  $y_D$ couplings are viable. 
Here too, one can construct models with large down-quark
mixings, but we leave the phenomenology of such models
for future study.

The model of~\secref{modela} provides a concrete realization 
of~\eqref{eq:mfvlike} 
using a flavor symmetry, but the approximate equality of the
messenger couplings and the Yukwas can hold much more generally
whenever the messengers and the Higgses have the same properties
with respect to the underlying theory of flavor.

\subsection{New mass patterns} \label{sec:new}
It is also possible to construct models with additional symmetries,
under which $D$ and $H_D$ transform differently.
Most of the models we consider below are of this type.
In all of these, the term $X H_D D$ is forbidden by holomorphy,
so that the superpotential is again of the form~\eqref{eq:superpot}.
To illustrate the basic mechanism consider a one-generation toy model.
We impose a $U(1)$ symmetry  broken by a spurion $\epsilon$ of charge $-1$, 
with the following charges, 
\beq
\label{toy}
H_D~(-1), \ \ \ d^c~(1), \ \ \   e^c~(1),\ \ \  l~(n\geq0)\ , 
\eeq   
and all other fields neutral. 
The term $X H_D \bar{D}_1$ cannot appear while the usual Yukawas are allowed.
In this case, the coupling $y_L$ is smaller than $Y_L$ by the factor
$\epsilon$, and the phenomenology of this model is not very interesting 
because the deviations from GMSB masses would be smaller than those induced
by the Yukawas. In the models we construct below, however, 
the new $U(1)$ will be part
of a $U(1)\times U(1)$ flavor symmetry, with the second $U(1)$ factor
compensating for this suppression, and leading to some entries 
$\left({y_L}\right)_{ij}> \left({Y_L}\right)_{ij}$.

\section{Generation dependent slepton spectra with a flavor symmetry}
\label{sec:examples}

We will assume that the hierarchies of the SM fermion 
masses are explained by a broken flavor symmetry, which
we take to be U$(1)_1\times$U$(1)_2$, with each  U(1) factor
broken by a spurion $\lambda_{1,2}$ of charge $-1$,
and with $\lambda_1\sim \lambda_2=\lambda\sim 0.1-0.2$.

The models are then completely specified by choosing $U(1)_1\times U(1)_2$
charges for the different fields.
We always take $H_U$, as well as all the messengers apart from $D\equiv D_1$
to be neutral under this symmetry.
In addition, we choose the charges of $H_D$ as $(0,-1)$,
with the $-1$  motivated by the fact that we want to eliminate
$H_D$ couplings to the supersymmetry-breaking field $X$
as explained in~\secref{new}. In fact, the $U(1)_2$ factor plays the role
of the $U(1)$ symmetry of the toy model of that section.
The models thus differ from each other because of the charges
of the matter fields and the messenger $D$,
and we will discuss different options below.

Since the SM matter fields transform non-trivially
under the flavor symmetry, the structure of the new coupling matrices
$y_L$ is affected by this symmetry as well, with some entries
suppressed by powers of $\lambda$, so that the flavor-changing
contributions are potentially suppressed by powers of $\lambda$.

In order to estimate these contributions and to determine whether
the models are viable, it is useful to work in terms of the
quantities $\delta_{i\neq{j}}$~\cite{Gabbiani:1996hi},
which are the basic quantities constrained by bounds on flavor 
violation.
Since we will be interested in the phenomenological predictions of
the models, it is useful to work in the slepton-mass basis,
so that the slepton mass differences and mixings are transparent.
One can then write (see for 
example,~\cite{Nir:2007xn})\footnote{Neglecting LR mixings, 
which is a good approximation in the models below.},
\beq\label{eq:delta}
\delta^A_{ij} \equiv ~\frac{\Delta\tilde{M}^2_{Aji}}
{\tilde{M}_{Aji}^2} \, K^A_{ij}\ ,
\eeq
where $A=L$ ($A=R$) refers to the lepton doublets (singlets), 
\beqa
\Delta \tilde{M}^2_{Aji} &=& \tilde{M}^2_{Aj}-\tilde{M}^2_{Ai}\ ,\nonumber\\
\tilde{M}_{Aji} &=&  \left[\tilde{M}_{Aj}+\tilde{M}_{Ai} \right]/2\ ,
\eeqa
and where $M_{Ai}$ is the mass of the slepton $i$,
and $K^A$ is the mixing matrix of the electroweak gaugino 
couplings\footnote{With a slight
abuse of notation, we use the same indices to label lepton
and slepton states.}.
Clearly, the flavor-changing contributions can be small
if either the mass-splittings or the inter-generation mixings
are small, or both. The example below will interpolate between
these options.

It will be convenient for our purposes to parametrize the experimental bounds 
as powers of $\lambda$. The most stringent bounds 
are from~\cite{Brooks:1999pu,Aubert:2009tk}, and using the results
of~\cite{Ciuchini:2007ha}, we have 
\beqa\label{eq:delta_constraints}
\delta^L_{12} &\lesssim& \lambda^4\,,~~~~~\delta^L_{13} 
\lesssim \lambda-\lambda^2\,,
~~~~~\delta^L_{23} \lesssim \lambda\,,\nonumber \\
\delta^R_{12} &\lesssim& \lambda^2\,,~~~~~\delta^R_{13} \lesssim \lambda\,,
~~~~~~~~~~~~\delta^R_{23} \lesssim \lambda\,, \\
\delta^{LR}_{12} &\lesssim& \lambda^5\,,~~~~~\delta^{LR}_{13} \lesssim \lambda^2\,,
~~~~~~~~~\delta^{LR}_{23} \lesssim \lambda^2\nonumber
\,.
\eeqa

\subsection{MFV-like masses with potentially large mixings}\label{sec:modela}
Choosing $D$ and $H_D$ to have identical flavor charges
results in MFV-like masses, since $y_L$ and $Y_L$ are
equal up to $O(1)$ coefficients.
With the flavor charges given in Table~\ref{tab:tablea},
\begin{table}[h]
\centering
\renewcommand{\arraystretch}{1.25}
\begin{tabular}{|c||c|c|c||c|c|c||c|c|}
\hline
Superfield & $l_1$ & $l_2$ & $l_3$ & $e^c_1$ & $e^c_2$ & $e^c_3$ 
& $H_d$ & $D$ \\
\hline\hline
$U(1)_1$ & $\ $4$\ $& $\ $2$\ $ & $\ $0$\ $ & $\ $1$\ $ & $\ $1$\ $ & $\ $0$\ $
 & $\ $0$\ $ &  $\ $0$\ $ \\
\hline
$U(1)_2$ & $\ $0$\ $ & $\ $2$\ $ & 4 & 1& $-1$ & $-2$ & $-1$ & $-1$ \\
\hline
\end{tabular}
\caption{Flavor charges of MFV-like model.}
\label{tab:tablea}
\end{table}
the desired lepton masses are obtained, and
\beqa\label{eq:basicyuk}
y_L \sim Y_L \sim 
\left(\begin{array}{ccc}
\lambda^5 & 0 & 0 \\
\lambda^5 & \lambda^3 & 0 \\
\lambda^5 & \lambda^3 & \lambda \\
\end{array}\right)
~.
\eeqa
Here and in the following, the entries are determined
to leading order in $\lambda$ and up to $\mathcal{O}(1)$ 
coefficients. 

One then finds, setting all terms suppressed by more than
six powers of $\lambda$ to zero (we denote such terms by ''$\sz$''),
\beqa\label{eq:model_A_LL_blocks}
\widetilde{m}^2_{LL} \sim \frac{\Lambda^2}{128\pi^4} 
\Bigg[N_5\, G_L {\bf 1_{3 \times 3}} - \frac{3}{2}  G_1
\left(
\begin{array}{ccc}
  \sz & \sz & \sz \\
 \sz & \lambda ^{6} & \lambda ^{6} \\
 \sz & \lambda ^{6} & \lambda ^{2}
\end{array}
\right) \Bigg]
\eeqa
and
\beqa\label{eq:model_A_RR_blocks}
\widetilde{m}^2_{RR} \sim 
\frac{\Lambda^2}{128\pi^4} \Bigg[
N_5\, G_R {\bf 1_{3 \times 3}} - 3\, G_1
\left(\begin{array}{ccc}
 \sz & \sz & \lambda ^{6} \\
 \sz &  \lambda ^{6} & \lambda ^{4} \\
 \lambda ^{6} & \lambda ^{4} & \lambda ^{2}
\end{array}\right) \Bigg]
~,
\eeqa
where we defined the mass scale $\Lambda \equiv F / M$ and the 
dimensionless numbers
\beqa
G_{L} \equiv \frac{3}{4} g_2^4 + \frac{5}{3} g_Y^4\,,~~~~~
G_{R} \equiv  \frac{20}{3} g_Y^4\,,~~~~~
G_{1} \equiv g_2^2 + 4 g_Y^2
\,.
\eeqa
The first term of each mass matrix in~\eqref{eq:model_A_LL_blocks},
\eqref{eq:model_A_RR_blocks} is 
the ordinary GMSB result and the second term is the contribution due 
to the new messenger-matter couplings. 
Note that the signs of the diagonal entries in these new contributions
are known: The $O(1)$ numbers multiplying the powers of $\lambda$
on the diagonals are positive.
We also get
\beqa\label{eq:model_A_LR_block}
\widetilde{m}^2_{LR} \sim -\frac{\Lambda v_d}{16\pi^2} 
\left[3
\left(\begin{array}{ccc}
 \sz & \sz & \sz \\
 \sz &  \sz & \sz \\
 \sz & \lambda ^{5} & \lambda ^{3}
\end{array}\right) + 
 \frac{\mu}{\Lambda/16\pi^2} \tan\beta\,
\left(\begin{array}{ccc}
\lambda^5 & 0 & 0 \\
\lambda^5 & \lambda^3 & 0 \\
\lambda^5 & \lambda^3 & \lambda
\end{array}\right) \right]
~,
\eeqa
The second term of $\widetilde{m}^2_{LR}$ is the standard $\mu$-term 
contribution, while the first comes from the $A$-term,
and is sub-dominant even for $\tan\beta\sim1$. 

Since the mass splittings in this case are of the order
of the mass splittings in MFV models, the model automatically satisfies
all flavor constraints, with a selectron and smuon that are
practically degenerate. 
The stau mass is split from the other masses by $O(\lambda^2)$,
coming from the 3-3 entries of the LL, RR and LR blocks (the latter
appears in  minimal GMSB models too). A similar effect is induced
by the running from the messenger scale to the weak scale,
and is probably the dominant effect since it's log-enhanced.
This too is a feature of minimal GMSB models, so the stau splitting
here is the same as in GMSB models, and this holds in all of our
models.

However, unlike  MFV models in which the fermion mass matrix
and the slepton mass matrix are diagonal in the same basis, 
this model predicts  $O(\lambda^2)$ 
mixings of $\tilde{e}_R-\tilde{\mu}_R$ and $\tilde{\mu}_R-\tilde{\tau}_R$.
The former might not be observable because of the small selectron-smuon
mass splitting, but the latter may be within reach of LHC experiments.

As explained before, this model will necessarily
contain couplings of the $D$ messenger to down quarks,
so that down squarks receive generation-dependent corrections
as well. These are largest when the third generation is
involved, with
\beq\label{eq:downdiff}
\frac{\Delta\tilde{M}^2_{ij}} {\tilde{M}_{ji}^2}\alt \frac1{N_5}\, y_b^2\ , 
\eeq
where $y_b$ is the bottom Yukawa.
The most severe constraint on the models is~\cite{Hiller:2008sv}  
$\delta^{d,LL}_{13}\delta^{d,RR}_{13}\alt 5\cdot 10^{-5}$,
but this is satisfied for $\tan\beta\sim 1$ or for $N_5=3$
even for  $\tan\beta\sim 5$.

\subsection{Selectron splitting}\label{sec:modelb}
In order to obtain some large entries in $y_L$,
these entries must involve smaller powers of $\lambda$
compared to the relevant entry of $Y_L$.
It is easy to achieve this by taking the $U(1)_1$ charge of $D$ to
be smaller than the  $U(1)_1$ charge of $H_D$ (which we took to be zero).
Consider for example the flavor charges of Table~\ref{tab:tableb}.
\begin{table}[h]
\centering
\renewcommand{\arraystretch}{1.25}
\begin{tabular}{|c||c|c|c||c|c|c||c|c|}
\hline
Superfield & $l_1$ & $l_2$ & $l_3$ & $e^c_1$ & $e^c_2$ & $e^c_3$ 
& $H_d$ & $D$ \\
\hline\hline
$U(1)_1$ & $\ $4$\ $& $\ $2$\ $ & $\ $0$\ $ & $\ $1$\ $ & $\ $1$\ $ & $\ $0$\ $
 & $\ $0$\ $ &  $\ -5\ $ \\
\hline
$U(1)_2$ & $\ $0$\ $ & $\ $2$\ $ & 4 & 1& $-1$ & $-2$ & $-1$ & $0$ \\
\hline
\end{tabular}
\caption{Flavor charges for~\secref{modelb}.}
\label{tab:tableb}
\end{table}
The large negative charge of $D$ has two consequences for the slepton spectrum.
First, most of the entries of $y_L$ vanish due to holomorphy~\cite{Nir:1993mx},
with only the $1-1$ entry surviving. Second, this entry is rather
large.
Thus, only the first-generation fields, whose charges are largest 
so that their masses would be the most suppressed, 
couple to the messenger sector,
and the modification of the selectron mass is appreciable.
In addition, because of this large negative charge, it is easy to choose 
charges for the down quarks
so that $y_D$ vanishes identically.

The resulting lepton Yukawas are as in~\eqref{eq:basicyuk}
while the new couplings are given by,
\beqa
y_L \sim
\left(\begin{array}{ccc}
\lambda & 0 & 0 \\
0 & 0 & 0 \\
0 & 0 & 0 \\
\end{array}\right)
~.
\eeqa

The LL and RR blocks are then,
\beqa
\widetilde{m}^2_{LL} \sim \frac{\Lambda^2}{128\pi^4} 
\left[N_5\, G_L {\bf 1_{3 \times 3}} - \frac{3}{2}\, G_1\, 
\left(
\begin{array}{ccc}
 \lambda ^2 & \sz & \sz \\
 \sz & \sz &  \sz \\
 \sz & \sz & \sz
\end{array}
\right) \right]
~,
\eeqa
and
\beqa
\widetilde{m}^2_{RR} \sim \frac{\Lambda^2}{128\pi^4} 
\left[N_5\, G_R {\bf 1_{3 \times 3}} -  3 G_1\,
\left(\begin{array}{ccc}
 \lambda^2 & \sz & \sz \\
\sz & 0 & 0 \\
\sz & 0 & 0
\end{array}\right) \right]
~.
\eeqa
The $A$-terms are negligible in this model.

The slepton mixings in this case arise solely from the
lepton mass matrix, and are given by,
\beq\label{eq:mixings}
K^L_{12}\sim \lambda^4\,,\ \ K^L_{13}\sim \lambda^8\,,\ \ 
K^L_{23}\sim \lambda^4\,;\ \ 
K^R_{12}\sim \lambda^2\,,\ \ K^R_{13}\sim \lambda^4\,,\ \ 
K^R_{23}\sim \lambda^2\,.
\eeq
The only significant $\delta$ 
is $\delta_{RR,12}\sim \lambda^4/N_5$ which is below
the bound.
In both the L- and the R-sectors, the selectron is lighter than 
the smuon by $\delta{m}\sim\lambda^2$.
Given that our estimates are parametric only, it is impossible 
to tell in these models whether the selectron
is the lightest slepton, since the stau masses are also driven
lower by $O(\lambda^2)$ both by running effects and by the
$\mu$ term contribution (the RGE contribution could be bigger for
a high messenger scale because it is logarithmically enhanced). 
In any case, the resulting spectrum
is very interesting, with the smuon being the heaviest slepton
and the selectron and stau lighter than the smuon, with
mass splittings around a few GeV
or even 10~GeV, and with $e-\mu$ and $\mu-\tau$ mixings
of a few percent in the R sector.

\subsection{Large mixings}\label{sec:modelc}
The previous model leads to small mixings
of the selectron with the other sleptons.
We can also obtain large selectron mixings by choosing
charges so that the new couplings are similar for the
three generations. In this case, of course, the mass
splittings are more constrained, so we want the size
of the new couplings to  be sufficiently small,
motivating the choice of charges for $D$ as shown
in Table~\ref{tab:tablec}.
\begin{table}[h]
\centering
\renewcommand{\arraystretch}{1.25}
\begin{tabular}{|c||c|c|c||c|c|c||c|c|}
\hline
Superfield & $l_1$ & $l_2$ & $l_3$ & $e^c_1$ & $e^c_2$ & $e^c_3$ 
& $H_d$ & $D$ \\
\hline\hline
$U(1)_1$ & $\ $2$\ $& $\ $2$\ $ & $\ $2$\ $ & $\ $4$\ $ & $\ $2$\ $ & $\ $0$\ $
 & $\ $0$\ $ &  $\ $0$\ $ \\
\hline
$U(1)_2$ & $\ $0$\ $ & $\ $0$\ $ & 0 & 1& $1$ & $1$ & $-1$ & 0 \\
\hline
\end{tabular}
\caption{Flavor charges for \secref{modelc}.}
\label{tab:tablec}
\end{table}

This yields an ordinary lepton Yukawa matrix of
\beqa
Y_L \sim
\left(\begin{array}{ccc}
\lambda^6 & \lambda^4 & \lambda^2 \\
\lambda^6 & \lambda^4 & \lambda^2 \\
\lambda^6 & \lambda^4 & \lambda^2
\end{array}\right)
\eeqa
which requires a somewhat small $\tan\beta$, 
and
\beqa
y_L \sim
\left(\begin{array}{ccc}
\lambda^7 & \lambda^5 & \lambda^3 \\
\lambda^7 & \lambda^5 & \lambda^3 \\
\lambda^7 & \lambda^5 & \lambda^3 \\
\end{array}\right)
~.
\eeqa
The resulting slepton mass matrices are then
\beqa
\widetilde{m}^2_{LL} \sim
\frac{\Lambda^2}{128\pi^4} \left[N_5\, G_L {\bf 1_{3 \times 3}} - \frac{3}{2} G_1
\left(
\begin{array}{ccc}
 \lambda ^6 & \lambda ^6 & \lambda ^6 \\
 \lambda ^6 & \lambda ^6 & \lambda ^6 \\
 \lambda ^6 & \lambda ^6 & \lambda ^6
\end{array}
\right) \right]
~,
\eeqa
and
\beqa
\widetilde{m}^2_{RR} \sim
 \frac{\Lambda^2}{128\pi^4} \left[N_5\,G_R {\bf 1_{3 \times 3}} - 3 G_1
\left(
\begin{array}{ccc}
 \sz & \sz & \sz \\
 \sz & \sz & \sz \\
 \sz & \sz & \lambda^{6}
\end{array}
\right) \right]
~.
\eeqa
The A-terms are negligible so that the only contribution to
the LR term is the usual $\mu$ term contribution.
The slepton masses are approximately degenerate in this case,
apart from the stau. 
The mixings of the R-sector are as in~\eqref{eq:mixings},
but the L-sector has $O(1)$ mixings, 
\beq
K^L_{12},\ K^L_{13},\  K^L_{23}\sim O(1)\ . 
\eeq

Finally, let us comment on the down sector in this model.
With the choice of $D$ charges as in Table~\ref{tab:tablec},
the down-messenger couplings would generically satisfy
\beq
y_{D, ij} \sim \lambda Y_{D, ij} \ .
\eeq 
Thus, the relative mass splittings in this case are
generically $O(\lambda^2)$ smaller than those of~\eqref{eq:downdiff},
and the models are consistent with flavor bounds
involving down squarks.

\subsection{Some large splittings and some large mixings}
\label{sec:modeld}
Finally, we present an example in which the $\tilde{e}_L$ and
the $\tilde{\mu}_R$ masses receive significant corrections, 
with a large $2-3$ mixing in the L-sector.
The flavor charges are given in Table~\ref{tab:tabled}.
\begin{table}[h]
\centering
\renewcommand{\arraystretch}{1.25}
\begin{tabular}{|c||c|c|c||c|c|c||c|c|}
\hline
Superfield & $l_1$ & $l_2$ & $l_3$ & $e^c_1$ & $e^c_2$ & $e^c_3$ 
& $H_d$ & $D$ \\
\hline\hline
$U(1)_1$ & $\ $2$\ $& $\ $2$\ $ & $\ $2$\ $ & $\ $2$\ $ & $\ $2$\ $ & $\ $0$\ $
 & $\ $0$\ $ &  $\ -4\ $ \\
\hline
$U(1)_2$ & $\ $0$\ $ & $\ $2$\ $ & 2& 2& $0$ & $0$ & $-1$ & 0 \\
\hline
\end{tabular}
\caption{Flavor charges for \secref{modeld}.}
\label{tab:tabled}
\end{table}
The lepton Yukawa matrix is as in~\eqref{eq:basicyuk},
and the messenger-lepton Yukawa couplings are
\beqa
y_L \sim 
\left(\begin{array}{ccc}
\lambda^2 & 1 & 0 \\
0 & 0 & 0 \\
0 & 0 & 0 \\
\end{array}\right)
~.
\eeqa
Just as in the model of~\secref{modelb}, the large and negative charge
of $D$ results in a large effect on the second generation,
with no effect on the third generation.
Furthermore, the messenger couplings to down quarks
will also vanish generically, since the total powers 
of $\lambda$ that should enter the down mass matrix entries, 
and therefore the total effective charge of these fields,
are typically smaller than those associated with the leptons.

These new couplings lead to
\beqa
\widetilde{m}^2_{LL} \sim
 \frac{\Lambda^2}{128\pi^4} \left[G_L {\bf 1_{3 \times 3}} 
- \frac{3}{2} G_1\, \left(
\begin{array}{ccc}
  1 & 0 &0 \\
  0 & 0 & 0 \\
  0 & 0 & 0
\end{array}
\right)
+  \left(
\begin{array}{ccc}
  1& \sz & \sz \\
  \sz & \lambda ^6 & \lambda ^6 \\
  \sz & \lambda ^6 & \lambda ^6
\end{array}
\right)
 \right]
~,
\eeqa
and
\beqa
\widetilde{m}^2_{RR} \sim 
\frac{\Lambda^2}{128\pi^4} \left[N_5\, G_R {\bf 1_{3 \times 3}} 
-3G_1\left(
\begin{array}{ccc}
 \lambda ^4  & \lambda ^2  &  0  \\
 \lambda ^2 & 1 &  0  \\
  0 &  0  & 0
\end{array}
\right) 
+\left(
\begin{array}{ccc}
 \lambda ^4  & \lambda ^2  &  \lambda ^6  \\
 \lambda ^2 &1 &  \lambda ^4  \\
  \lambda ^6  & \lambda ^4  & 0
\end{array}
\right) 
\right]
~.
\eeqa
The A-terms are again very small, with
\beqa
\widetilde{m}^2_{LR} \sim 
-\frac{\Lambda v_d}{16\pi^2}
\left[
\left(
\begin{array}{ccc}
 \lambda ^5 & \sz & 0 \\
  \lambda ^5 &  \lambda ^3 & 0 \\
  \lambda ^5 &  \lambda ^3 & 0
\end{array}
\right) + 
\frac{\mu}{\Lambda/16\pi^2} \tan\beta\,
\left(\begin{array}{ccc}
\lambda^5 & 0 & 0 \\
\lambda^5 & \lambda^3 & \lambda \\
\lambda^5 & \lambda^3 & \lambda
\end{array}\right) \right]
~.
\eeqa
The RR mixings are as in~\eqref{eq:mixings}, and the 
LL mixing are negligible apart from 
$K^L_{23}= O(1)$. 
The constrained quantities~\eqref{eq:delta_constraints} for the LL block 
are negligible. 
In the RR block, $\delta_{RR,12}$ and
$\delta_{RR,23}$ are of order $\lambda^{2} / N_5$, 
saturating the bound on $\delta_{RR,12}$ for small $N_5$. 
The same holds for  
$\delta_{LR,12} \sim \lambda^{5} / N_5$. 
The other $\delta_{LR}$'s are negligible. 
Since the model is only specified up to $O(1)$ parameters,
we see that it can be consistent with bounds on flavor-violation
for parts of the parameter space.

This model has a very interesting spectrum.
The $\tilde{e}_L$   has a large mass splitting compared
to the other L-sleptons, 
\beq
\frac{\Delta\tilde{M}^2_{L1i}}{\tilde{M}_{L}^2}\sim \frac1{N_5}\,,\ \  i=2,3\,,
\eeq
and hardly mixes with the $\tilde{\mu}_L$,  $\tilde{\tau}_L$.
In addition, the $\tilde{\mu}_L-\tilde{\tau}_L$ mixing is large.

In the R-sector,
\beq
\frac{\Delta\tilde{M}^2_{R2i}}{\tilde{M}_{L}^2}\sim \frac1{N_5}\,, \ \ i=1,3
\eeq 
so that the R-smuon is significantly
split from the other R-sleptons.
Since, in addition, the masses of
$\widetilde{e}_R$ and the staus have $O(\lambda^2)$ corrections
to their GMSB masses, all six sleptons are separated in mass.

\section{Conclusions}
\label{sec:conclusions}
We presented models in which slepton masses are generated
by messenger fields, through gauge and superpotential interactions.
If such spectra are measured at the LHC, the GMSB structure will
be apparent in the gaugino spectrum, with the slepton masses
clearly indicating some flavor-dependent mediation of supersymmetry
breaking, and providing additional handles on the source
of fermion masses in the standard model.
We concentrated on slepton masses, but as we explained, it is
straightforward to generalize this construction to include
messenger couplings to squarks. 

It would also be interesting to examine mechanisms for
generating the mu term in these models, since the flavor
symmetries we discussed often forbid this term.
The models may also accommodate large couplings
of the Higgs to the supersymmetry-breaking sector in the 
spirit of~\cite{Csaki:2008sr}.

Finally, while our models are based on flavor symmetries,
it would be interesting to consider alternative
frameworks for controlling both
the Yukawa couplings and the matter-messenger couplings.


\section*{Acknowledgments}

We thank Yossi Nir, Yuval Grossman, and Yuri Shirman for helpful discussions.  
Parts of this work were done during a sabbatical
year of YS at UCI. YS thanks the UCI hep group for its warm hospitality
and financial support during this year.
The work of YS and PZS was supported by the
Israel Science Foundation (ISF) under Grant No.~1155/07,
and by the United States-Israel Binational Science
Foundation (BSF) under Grant No.~2006071.


\providecommand{\href}[2]{#2}\begingroup\raggedright\endgroup

\end{document}